\documentclass[apj]{emulateapj}
\usepackage{mathptmx}

\newcommand  \kms      {\ifmmode {\rm km\,s}^{-1} \else km\,s$^{-1}$\fi}

\newcommand  \ergs     {\ifmmode {\rm ergs\,s}^{-1} \else ergs s$^{-1}$\fi}
\newcommand  \ergcms   {\ifmmode {\rm ergs\,cm}^{-2}\,{\rm s}^{-1}
                        \else ergs\,cm$^{-2}$\,s$^{-1}$\fi}
\newcommand  \ergcmsA {\ifmmode{\rm ergs\,cm}^{-2}\,{\rm s}^{-1}\,{\rm\AA}^{-1}
                        \else ergs\,cm$^{-2}$\,s$^{-1}$\,\AA$^{-1}$\fi}
\newcommand \ergcmsHz {\ifmmode{\rm ergs\,cm}^{-2}\,{\rm s}^{-1}\,{\rm Hz}^{-1}
                        \else ergs\,cm$^{-2}$\,s$^{-1}$\,Hz$^{-1}$\fi}
\newcommand  \phcms    {\ifmmode {\rm ph\,cm}^{-2}\,{\rm s}^{-1}
                        \else ,ph\,cm$^{-2}$\,s$^{-1}$\fi}
\newcommand  \phcmsA   {\ifmmode {\rm ph\,cm}^{-2}\,{\rm s}^{-1}\,{\rm\AA}^{-1}
                        \else ph\,cm$^{-2}$\,s$^{-1}$\,\AA$^{-1}$\fi}
\def\micron{\ifmmode \mu{\rm m} \else $\mu$m\fi}
\def\kms{\ifmmode {\rm km\,s}^{-1} \else km\,s$^{-1}$\fi}
\def\Hubble{\ifmmode {\rm km\,s}^{-1}\,{\rm Mpc}^{-1}
        \else km\,s$^{-1}$\,Mpc$^{-1}$\fi}
\def\ergsec{\ifmmode {\rm ergs\;s}^{-1} \else ergs s$^{-1}$\fi}
\def\ergscm{\ifmmode {\rm ergs\,s}^{-1}\,{\rm cm}^{-2}
          \else ergs\,s$^{-1}$\,cm$^{-2}$\fi}
\def\ergscmA{\ifmmode {\rm ergs\,s}^{-1}\,{\rm cm}^{-2}\,{\rm \AA}^{-1}
          \else ergs\,s$^{-1}$\,cm$^{-2}$\,\AA$^{-1}$\fi}
\def\ergscmHz{\ifmmode {\rm ergs\,s}^{-1}\,{\rm cm}^{-2}\,{\rm Hz}^{-1}
          \else ergs\,s$^{-1}$\,cm$^{-2}$\,Hz$^{-1}$\fi}
\def\Msun{\ifmmode M_{\odot} \else $M_{\odot}$\fi}
\def\Lsun{\ifmmode L_{\odot} \else $L_{\odot}$\fi}
\def\qo{\ifmmode q_{0} \else $q_{0}$\fi}
\def\Ho{\ifmmode H_{0} \else $H_{0}$\fi}
\def\ho{\ifmmode h_{0} \else $h_{0}$\fi}
\def\qo{\ifmmode q_{0} \else $q_{0}$\fi}
\def\ao{\ifmmode a_{0} \else $a_{0}$\fi}
\def\to{\ifmmode t_{0} \else $t_{0}$\fi}
\def\gtsim{\raisebox{-.5ex}{$\;\stackrel{>}{\sim}\;$}}
\def\Halpha{\ifmmode {\rm H}\alpha \else H$\alpha$\fi}
\def\Hbeta{\ifmmode {\rm H}\beta \else H$\beta$\fi}
\def\hb{\ifmmode {\rm H}\beta \else H$\beta$\fi}
\def\Hgamma{\ifmmode {\rm H}\gamma \else H$\gamma$\fi}
\def\Hdelta{\ifmmode {\rm H}\delta \else H$\delta$\fi}
\def\Lya{\ifmmode {\rm Ly}\alpha \else Ly$\alpha$\fi}
\def\Lyb{\ifmmode {\rm Ly}\beta \else Ly$\beta$\fi}
\def\hi{\ifmmode \mbox{{\rm H}\,{\sc i}} \else H\,{\sc i}\fi}

\def\heiiop{He\,{\sc ii}\,$\lambda4686$}

\def\ciii{\ifmmode {\rm C}\,{\sc iii} \else C\,{\sc iii}\fi}
\def\civ{C\,{\sc iv}\,$\lambda1549$}

\def\nv{N\,{\sc v}\,$\lambda1240$}

\def\oiii{[O\,{\sc iii}]\,$\lambda5007$}

\def\o5007{[O\,{\sc iii}]\,$\lambda5007$}

\def\ne212m {[Ne\,{\sc ii}]\,$12.8 \mu m$}

\def \Lop{L$_{5100}$}
\def \Ledd{$L/L_{Edd}$}

\def  \kms         {\hbox{km s$^{-1}$}}          
\def  \ergs        {\hbox{erg s$^{-1}$}}              

\def  \La          {\ifmmode {\rm Ly}\alpha \else Ly$\alpha$\fi}
\def  \Ka          {\ifmmode {\rm K}\alpha \else K$\alpha$\fi}
\def  \Lb          {\ifmmode {\rm L}\beta \else L$\beta$\fi}
\def  \Ha          {\ifmmode {\rm H}\alpha \else H$\alpha$\fi}
\def  \Hb          {\ifmmode {\rm H}\beta \else H$\beta$\fi}
\def  \Pa          {\ifmmode {\rm P}\alpha \else P$\alpha$\fi}
\def  \CIIIb       {\ifmmode {\rm C}\,{\sc iii]}\,\lambda1909
                     \else C\,{\sc iii]}\,$\lambda1909$\fi}
\def  \CIV         {\ifmmode {\rm C}\,{\sc iv}\,\lambda1549
                     \else C\,{\sc iv}\,$\lambda1549$\fi}
\def  \MgII         {\ifmmode {\rm Mg}\,{\sc ii}\,\lambda2798
                     \else Mg\,{\sc ii}\,$\lambda2798$\fi}
\def  \OVI         {\ifmmode {\rm O}\,{\sc vi}\,\lambda1035
x
                     \else O\,{\sc vi}\,$\lambda1035$\fi}

\shorttitle{Mass accretion rate and metalicity in AGNs}
\shortauthors{Netzer and Trakhtenbrot}

\begin{document}

\title{Cosmic evolution of mass accretion rate and metalicity in \\ active galactic nuclei}

\author{
Hagai Netzer,\altaffilmark{1,2}
Benny Trakhtenbrot\altaffilmark{1}
}

\altaffiltext{1}
          {School of Physics and Astronomy and the Wise
                Observatory, The Raymond and Beverly Sackler Faculty of
                Exact Sciences, Tel-Aviv University, Tel-Aviv 69978,
                Israel}
\altaffiltext{2}
 {Max-Planck-Institut f\"ur extraterrestrische Physik, Postfach 1312, 85741 Garching, Germany}

\begin{abstract}
We present line and continuum measurements for
9818 SDSS type-I active galactic nuclei (AGNs) with $z \le 0.75$.
The data are used to study the four dimensional space of black hole mass,
normalized accretion rate (\Ledd), metalicity and  redshift.
The main results are:
1. \Ledd\ is smaller for larger mass black holes at all redshifts.
2. For a given black hole mass,
\Ledd$\propto z^{\gamma}$ or $(1+z)^{\delta}$ where the
slope $\gamma$ increases with black hole mass. The mean slope is
similar to the star formation rate slope over the same redshift interval.
3. The FeII/\hb\ line ratio is significantly correlated with \Ledd. It also shows
a weaker negative dependence on redshift.
Combined with the known dependence of metalicity on accretion rate,
 we suggest that the  FeII/\hb\ line ratio is a metalicity indicator.
4. Given the measured accretion rates, the growth times of most AGNs exceed the age of the
universe. This suggests past episodes of faster growth for all those sources.
Combined with the FeII/\hb\ result, we conclude that the broad emission lines metalicity goes through
cycles and is not a monotonously decreasing function of redshift.
5. FWHM(\oiii) is a poor proxy  of $\sigma_*$  especially for high \Ledd.
6. We define a group of narrow line type-I AGNs (NLAGN1s) by their luminosity (or mass) dependent
\hb\ line width. Such objects have \Ledd$\ge 0.25$ and they comprise 8\% of the type-I population.
Other interesting results include negative Baldwin relationships
for EW(\hb) and EW(FeII) and a relative increase of the red part of the \hb\ line
with luminosity.

\end{abstract}


\section{Introduction}
Recent progress in reverberation mapping of type-I (broad emission lines) active galactic
nuclei (AGNs) provide a reliable method for deriving the size of the broad line region (BLR)
as a function of the optical continuum luminosity, the \hb\ line luminosity, the
ultraviolet continuum luminosity and the X-ray (2--10 keV) luminosity (Kaspi et al. 2005, hereafter K05
and references therein). This has been used in numerous papers to obtain a
 ``single epoch'' estimate of the black hole (BH) mass
 by combining the derived BLR size with a measure of the gas
 velocity obtained from emission line widths, e.g. FWHM(\hb)
  (e.g. Netzer 2003; Vestergaard 2004; Shemmer et al. 2004, hereafter S04; Baskin and
Laor 2005).
There are obvious limitations to such methods due to the scatter in the BLR size, the source luminosity
and  line width. These translate to a factor of 2--3 uncertainty on
the derived masses when using the \hb\ line width as the velocity measure and the optical continuum
luminosity ($\lambda L_{\lambda}$ at 5100\AA, hereafter \Lop) for estimating the BLR size.
  The uncertainty is larger, and
perhaps systematic, when other emission lines are used.
Additional uncertainties are associated with the
currently limited luminosity range,
$10^{42}\le$\Lop$\le 10^{46}$ \ergs, 
thus mass determination for very luminous
sources must be based on extrapolation.
Notwithstanding the uncertainties, such methods can provide BH mass and accretion rate in large
AGN samples that cannot be obtained in any other way.

This paper follows several earlier studies of type-I AGNs in the Sloan Digital Sky Survey 
(SDSS; see York et al. 2000) spectroscopic archive.
The main  aims are to obtain BH masses and accretion rates for a large number
of such sources and to understand the time evolution and the metal content of the BLR gas.
Earlier studies of this type include, among others, McLure  and Jarvis
(2002) and McLure and Dunlope (2004). Here we use the latest SDSS public data archive and
 a much larger number of sources compared with earlier studies.
 This enables us to obtain statistically significant
correlations for various sub-samples of the data.
 We also introduce new, more robust methods for measuring the emission lines thus enabling
 more reliable study of metalicity and accretion rate.

This paper is organized as follows. In \S2 we introduce our sample and
explain the method of analysis and the various measurements performed on the SDSS spectra. In \S3
we show the more important correlations pertaining to type-I sources and in \S4 we give a
discussion of the main results, including the
derived redshift dependent accretion rate which we compare with the known star formation rate
 over the same redshift interval.

\section{Basic measurements}

\subsection{Sample selection}

The present sample contains all broad line AGNs in the SDSS sample from data release 4 
(DR4, see Adelman-McCarthy et al. 2006 and references therein)
with a redshift cut at $z = 0.75$. This more than doubles the number of such sources
in DR3 (Schneieder et al. 2005). 
The sources are defined as QSOs by the target selection
algorithm but will be addressed here as type-I AGNs.
The limit of 0.75 is dictated by the wavelength band of the
 SDSS spectroscopy and is designed
to allow the measurement of the \hb\ and \oiii\ lines and the 5100\AA\ continuum
 in all sources. The inclusion of \oiii\ is
required for different line velocities and for omitting very few sources with such weak broad lines that they
are confused with type-II AGNs.
The requirement on \hb\ is motivated by the work of Kaspi et al. (2000) and K05
which present the only available reverberation mapping sample. This low redshift sample includes
\hb\ and \Lop\ measurements for 35 sources and much fewer UV line and continuum measurements.
The BLR size is obtained from the measured \hb\ lag (or a combination of $H\alpha$ and \hb\ lags)
and the velocity from the observed FWHM(\hb).
Different lines have been used in several other papers (e.g. Vestergaard 2004; Vestergaard and Peterson 2006) yet
their use is more problematic since it requires a secondary calibration against \hb\ and \Lop.
Of the two UV lines most commonly used, \MgII\ is probably more reliable (e.g. McLure and Jarvis 2002;
McLure and Dunlop 2004) and
\civ\ suffers from various systematic effects (Baskin and Laor 2005 and references therein).
We limit the present study to the use of \hb\ and defer
 the study of higher redshift AGNs, based on the \MgII\ line, to a later publication.

The SDSS archive (Stoughton et al. 2002) was queried for QSOs with $z \le 0.75$.
This returned 14554 objects out of which 1212 were found to have
problematic mask data (spectra with more than 30\% flagged pixels;
see Stoughton et al.  table 11 for details). These objects are not included in the
analysis.  We decided to stick with the original magnitude limit of the sample at the 
time of target selection.  This means that QSOs which are
fainter than $i(mag)$=19.1 in the TARGET database, but were later added to the BEST database due to 
improved photometry, or objects that were added to the QSO database after their spectra were 
examined as different science targets (thus listed as ${\rm TARGET\_ QSO\_ FAINT}$ by the SDSS target
selection algorithm; see \S4.8.4 and table 27 in Stoughton et al.) were not included.
This removes 1927 additional sources from the sample leaving
11415 sources with $i(mag) \le 19.1$.

The spectra of all sources were corrected for Galactic extinction using the dust maps of Schlegel et al. (1998)
and the Cardelli et al. (1989) model. Next, pixels with problematic MASK values (as defined above) are 
replaced by local interpolation and the spectra are de-redshifted
and put on a uniform wavelength scale of 1\AA\ per pixel.

\subsection{Line and continuum measurements}

Our work make use of the SDSS spectra as obtained from the archive without any attempt to subtract
the galaxy contribution. This introduces some uncertainties at the low luminosity
end of the sample and is addressed below in those cases where we consider it to be important for the
derived correlations. The general issue of stellar contribution to AGN spectra of
various luminosities is discussed in Bentz et al. (2006).

The emission lines measured in this work
are \oiii, \hb\ and the FeII line complex between the wavelengths
of 4000 and 5300\AA. 
We first fit a linear continuum between two continuum bands: 4430--4440\AA\
and 5080--5120\AA. This continuum is subtracted from the spectrum and a five Gaussian model is
fitted to  the profiles of the two [O\,{\sc iii}] lines and \hb.
The [O\,{\sc iii}]$\lambda\lambda4959,5007$\AA\ lines are forced to have their theoretical intensity ratio (3.0),
the same FWHM and their known wavelengths. The \hb\ line is fitted with two broad Gaussians and a single
narrow Gaussian with FWHM and wavelength that agree with the [O\,{\sc iii}] lines to within
200 \kms.
At this stage, the  narrow \hb\ component is not allowed to exceed 0.7 the intensity of \oiii.
The broad \hb\ Gaussians are limited to FWHM of 1500--5000 \kms\ and a shift of up to 500 \kms\ relative
to \oiii. These first limits are necessary in order to avoid the inclusion of some neighboring FeII
lines, mostly on the blue side of the \hb\ profile, and to properly model the broad \hb\ components.

The next step involves fitting and deblending of FeII lines.
The Boroson \& Green (1992; hereafter BG92) Fe II emission
template was convolved with a single Gaussian, which has {\it the same FWHM} as the broad  \hb\
from the previous fit. The resulting FeII model is fitted to the
spectra over the  region 4400--4650\AA. This region was chosen
to minimize the effect of contamination with \hb, HeII$\lambda 4686$ and \oiii. Next we measure the
contribution of the FeII lines  to the previously defined continuum bands, subtract this and refit the
continuum. The entire procedure is repeated and the newly defined FeII template is subtracted from
the continuum-subtracted spectrum. This leaves us with a FeII-free spectrum.

The final stage involves, again, a five Gaussian fit to the [O\,{\sc iii}] and \hb\ lines.
At this stage we relax the conditions and allow a broader range of Gaussians (as
narrow as the \oiii\ line and as broad as 20,000  \kms), and
a larger shift between the narrow \hb\ and \oiii\ (450 \kms).
The 20,000 \kms\ limit is required since almost all cases exceeding this limit are found
to show poor quality fits (see below).
We also allow shifts of up to $\pm 1000$ \kms\ between each of the two broad \hb\ components
and the wavelength of the narrow \hb\ line.
All line and continuum measurements used later in the statistical analysis
 are based on the final stage of the fit and the second FeII fit.
We found this multi-stage method to be more robust, and less prone to fitting inaccuracies compared
with a single-stage fit of all line and components.

Regarding FWHM(\oiii), we use an ``intrinsic''  line width which is obtained by
properly subtracting a uniform instrumental line width of FWHM=140 \kms.
We have decided not to include in the fit the blue wing commonly observed in the line profile.
This removes a potential complication due to blending with the broad
\hb\ profile and is also in line with earlier measures of the line in type-I and type-II sources. 
As explained in Greene \& Ho (2005), a single component \oiii\ line 
is a good indicator of $\sigma_*$. This point is further discussed in \S3.5.
We do not include in the analysis the \heiiop\ line which is clearly seen in
many of the sources.

As for the broad \hb\ line, we measure its rest equivalent width (EW(\hb)) and FWHM(\hb) and
also divide the combined line profile into two parts, at the rest wavelength of 4862\AA\ (as determined by
the wavelength of \oiii). The line flux is then measured in each part.
We note that some earlier works followed BG92 and use EW(FeII) as a  replacement for the iron line intensities.
Here we define L(FeII) by integrating the line flux over
the wavelength range of 4434--4684\AA\ (e.g. S04). In what follows FeII/\hb\ stands for  L(FeII)/L(\hb).

The automated procedure to measure line and continuum fluxes suffers from various problems
and limitations, some
that are difficult to cure. To allow better control of the fit results, we make
use of the $R^2$ and $\chi^2$ statistics (both per degree of freedom).
 These are combined in various ways to reject sources
that do not fall within the predefined limits of $R^2 > 0.2$ and $\chi^2<5$.
A common problem is encountered in weak flux sources with very large FWHM(\hb). 
In many of these cases, mostly those at small redshift, the stellar contribution
to \Lop\ is of the same order as the non-stellar continuum. The poor signal-to-noise
(S/N) in some of those makes it difficult to properly fit the low contrast \hb\ profile. A similar
problem occurs with very broad low contrast FeII blends which can result in erroneous, very large
FeII/\hb. Such objects are usually found and rejected by their small value of $R^2$.
Given the strong dependence of luminosity on redshift, it is not surprising that such problems
are less common at higher redshifts. Thus, the fit quality at high redshift is usually better even
in sources that have low S/N due to being close to the flux limit of the sample. Our
rather conservative approach results in the rejection of about
13\% of all sources due to poor fits. Relaxing those criteria to include about half of the rejected
sources in the analysis change very little the conclusions of the paper.

\begin{figure}
\plotone{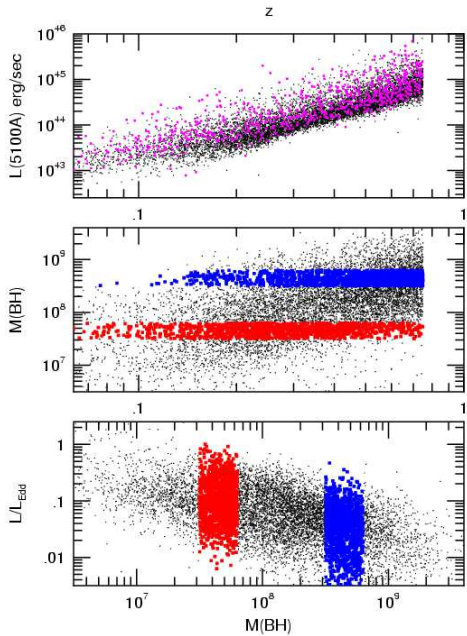}
\caption{
Basic sample properties. {\it Top panel:} \Lop\ vs. redshift for RL (magenta points)
and RQ (black points) type-I AGNs.
{\it Middle panel:}
M(BH) vs. redshift for type-I RQ-AGNs.
{\it Bottom panel:} \Ledd\ vs. M(BH) for type-I RQ-AGNs.
Here and in all other diagrams, small black points mark the entire RQ sample while the
colored dots represent sources
in two mass groups: $10^{7.5-7.8}$ \Msun\ (red) and $10^{8.5-8.8}$ \Msun\ (blue).
}
\label{fig:M_z_mdot}
\end{figure}

Radio information is readily available from the SDSS archive that
provides FIRST (First Images of the Radio Sky at Twenty centimeters)
data for all sources (e.g. Richards et al. 2002; McLure and Jarvis 2004).
This enables us to divide the sample into radio loud (RL) and
radio quiet (RQ) AGNs.
The final sample that it analyzed here includes 9818 sources out of which 812 are RL-AGNs.
Its luminosity distribution as a function of redshift is shown in Fig.\ref{fig:M_z_mdot}. 

\subsection{Luminosity black hole mass and accretion rate}

The combination of \Lop\ and FWHM(\hb) was converted to BH mass using the most recent
reverberation mapping results of K05. This gives
\begin{equation}
M(BH)= 1.05 \times 10^8 \left [ \frac {L_{5100}}{10^{46}\, {\rm erg\,sec^{-1}}} \right] ^{0.65}
     \left [ \frac {FWHM(H_{\beta})}{1000 \,km/sec} \right] ^2 \,\, \Msun \, .
\label{eq:M_L}
\end{equation}
The slope chosen here (0.65) is obtained from the R$_{\rm BLR}$--\Lop\ 
relationship of K05.
More recent papers (Vestergaard and Peterson 2006; Bentz et al. 2006)
suggest a flatter slope (0.52) based on improved
host galaxy subtraction in several of the low luminosity sources in the K05 sample,
and the omission of several sources from the original list.
We note that the inclusion of all the K05 sources in the Bentz et al. (2006) analysis,
using the BCES method (see the Bentz et al. paper
for explanation and their Table 1 for results)
gives a slope of $0.57 \pm 0.08$ which is entirely consistent with the K05
slope of $0.66\pm0.07$.  Moreover, there are other reasons for choosing a slope steeper than 0.52, e.g. 
using L(\hb) as a proxy for the UV continuum luminosity (which is not affected by
the host galaxy subtraction) gives such a slope. Finally, removing several of the faintest sources from
the K05 sample also justifies this slope. For example, in 91\% of the sources in our sample  \Lop$>10^{43.5}$ \ergs.
Thus it is justified not to include the K05 sources with \Lop$<10^{43.5}$ \ergs\  when searching for the 
most suitable reverberation mapping relationship for the SDSS sample (this
also avoids complications due to host galaxy contamination). This smaller sample produces a relationship very similar to the
one given by eqn.~\ref{eq:M_L}. 

The normalized accretion rate, \Ledd, is given by 
\begin{equation}
L/L_{Edd} = \frac{f_L L_{5100} }{1.5 \times 10^{38} [M(BH)/\Msun] } \,\, ,
\label{eq:Mdot}
\end{equation}
where $f_L$ is the bolometric correction factor. Estimates of  $f_L$
requires further assumptions about the spectral
energy distribution (SED), in particular the luminosity of the unobserved, extreme UV continuum.
This issue has been discussed in great detail in numerous publication. Estimates 
for  $f_L$  in type-I AGN range between 5 and about 13
(e.g. Elvis et al. 1994; Kaspi et al. 2000; Netzer 2003; Marconi et al. 2004; S04).
Excluding dust reprocessed radiation (that was included in the Elvis et al. 1994 SED),
the range is about 5--10.
There are clear indications from general considerations of the AGN luminosity function
(Marconi et al. 2004), as well as from direct spectroscopy (e.g. Scott et al. 2004; Shang et al. 2005), that
the shape of the SED  is luminosity dependent and the ratio is larger in
lower luminosity AGNs. 
Since the exact luminosity dependence is not known (see however Marconi et al. 2004, Fig. 3
for a specific suggestion) we adopt a constant ratio of $f_L=7$  which is roughly the mean
between the value deduced for weak and strong UV bump sources. A more advance analysis
can involve the actual Marconi et al. values. For the range of \Lop\ considered here
(Fig. 1), this would cover the range $6.2<f_L<8.3$.

The rest of this paper includes various diagrams involving mass, luminosity and accretion
rate of AGNs at different redshifts. Despite the specific expressions given in
eqns.~\ref{eq:M_L} and \ref{eq:Mdot}, none of those correlations is trivial since the 
dependence of luminosity on the BH mass is unknown a priori. Our work
can be considered as a study of the four dimensional distribution of type-I AGNs
along the coordinates of M(BH), $\dot{M}(BH)$, metalicity and redshift.

The SDSS is a flux limited sample and hence many of the derived correlations are affected by
its redshift dependence incompleteness.
In the following we take into account the flux limit of the sample which is transformed, as appropriate,
to the various correlations under study. The limits on \Lop, \Ledd\ etc. are
obtained from the (reddening corrected) $i(mag)$,
 the redshifts and a k-correction that assumes $L_{\nu} \propto \nu^{-0.5}$
across the wavelength range of interest (5100--8000\AA).

Fig.\ref{fig:M_z_mdot} shows  BH mass as a function of redshift for all RQ-AGNs in our sample. 
It confirms the well known tendency for finding  larger active black holes at higher
redshifts, a result that was already presented by McLure \& Dunlope (2004) over a wider redshift range.
The lower boundary at each redshift is a clear manifestation of the
strong correlation of M(BH) and \Lop\ in our flux limited sample.
The diagram is very similar to the one presented in McLure \& Dunlope (2004)
where a smaller sample and somewhat
different expressions for deriving M(BH) and \Ledd\ have been used.
Here, and in several other diagrams, we focus on two mass groups that are marked in two different colors:
$10^{7.5-7.8}$ \Msun\ (red, 1207 sources) and $10^{8.5-8.8}$ \Msun\ (blue, 1422 sources). 

McLure and Jarvis (2004) used a sample of about 6,000 SDSS QSOs to investigated the mass of the
central BHs as a function of radio properties.  They found that RL-QSOs host, on the
average, more massive BH for a given optical continuum luminosity. The mean M(BH) difference
between RL-AGNs and RQ-AGNs found by those authors is 0.16 dex. This was attributed to the larger mean FWHM(\hb) in
radio loud sources.  We checked these findings by comparing the luminosity, BH mass and \Ledd\ for the
two populations in our $z \le 0.75$ sample. 
Using the medians of all properties, we find a difference of 0.11 dex in \Lop, a factor of 1.06 in FWHM(\hb), and
a difference of 0.15 dex in M(BH) (the differences in the mean are somewhat larger). The difference
in M(BH) is, of course, the result of the differences in luminosity and line width between 
the two populations (see eq. \ref{eq:M_L}).

Fig.~\ref{fig:M_z_mdot} 
also shows \Ledd\ vs. M(BH) for the RQ-AGN sample as well as for the two mass groups.
The low accretion rate limit is determined, mostly,
by the flux limit of the sample. Thus, we expect that more object with small \Ledd\
are missing at higher redshifts.
The highest accretion rate sources in each mass group are not affected by this limit. Based of
these sources we 
confirms that the lowest mass BHs are the fastest accretors, a result that 
was already noted by McLure and Dunlop(2004).

\section{Results}

\subsection{Accretion rate vs. redshift}
\begin{figure}
\plotone{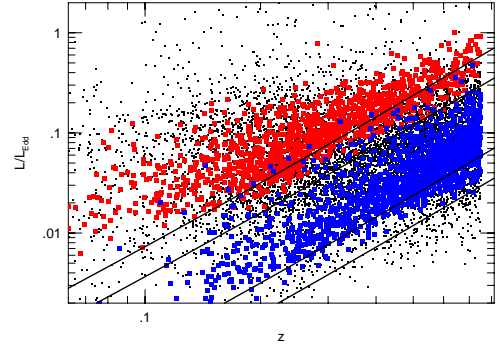}
\caption{\Ledd\ vs. redshift for our SDSS sample of type-I RQ-AGNs. The solid lines
represent the limiting accretion
rates obtained from the flux limits for the
two mass groups (see text). Symbols are the same as those of Fig.1.
}
\label{fig:accretion_z}
\end{figure}

The correlation of \Ledd\ with redshift for the entire sample, as well as for the two mass groups,
 is shown Fig.~\ref{fig:accretion_z}. The diagram shows also two computed lower limits
 for each of the two groups corresponding to the flux limits of the two mass boundaries.
The visual impression is somewhat
misleading since much of the apparent correlation is due to the  lower limit on \Ledd\ resulting from
the flux limits on the various mass groups.
The presence of such flux limits considerably complicate the analysis and
we have searched for rigorous ways to  derived the intrinsic z-\Ledd\ dependence.
Below we detail the two methods that were used, the peak of the \Ledd\ distribution
 and the Efron and Petrosian (1992) truncated permutation test.

\subsubsection{The peak distribution method}
\begin{figure}
\plotone{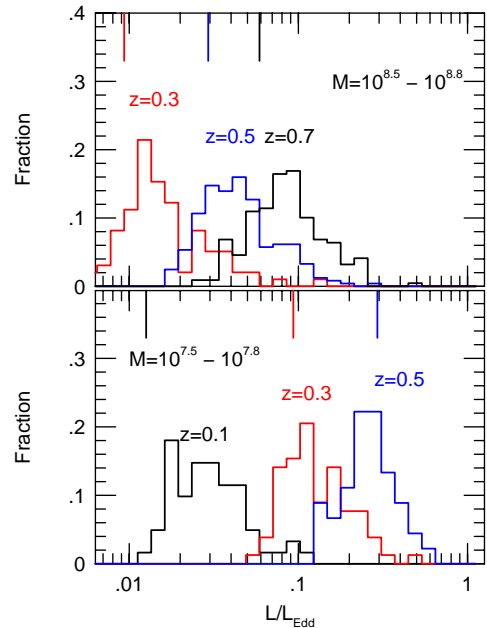}
\caption{\Ledd\ histograms for various redshift bands, as marked, for the low (bottom)
and high (top) mass groups. The redshift bins are defined in the text.
The vertical lines at the top of each panel mark the conservative
lower limit on \Ledd\ obtained from the flux limit of the same color histogram (see text for more details).
}
\label{fig:histogram_M4_M5_M6}
\end{figure}

  We first study the distribution of \Ledd\ in several redshifts bins that are just
  wide enough to include a reasonable number of sources. These histograms
  (see Fig.~\ref{fig:histogram_M4_M5_M6}) are used to define the peak and the envelope of the \Ledd\ vs. redshift
  relationship. The bins  
are defined such that $\Delta z /z=0.1$ and include from 50 to
230 sources, with an average number of about 120 sources per bin.

  Such a procedure  is only meaningful if the peak of the distribution is clearly defined, i.e.
  if the value of \Ledd\ corresponding to the maximum number of sources is {\it larger} than the accretion rate
  imposed by the flux limits of the sample.
  There are four such limits for each bin: two corresponding to the redshift boundaries and
  two to the mass boundaries. The largest of those four  (i.e. the most conservative
  limits) are shown in Fig.~\ref{fig:histogram_M4_M5_M6} for each of the two mass groups and several redshift bins.
  As evident from the diagram, the peaks of the \Ledd\ distribution in the lower mass group (bottom part of the
  diagram) are
  only meaningful for $z \le 0.3$. Beyond this redshift the method is not applicable since
 the SDSS survey is not deep enough to properly sample many sources and the limiting
 \Ledd\ is larger than the value at the peak.
The larger mass group (top panel) is less problematic and the
  histograms show that the peaks are well defined for $0.3 \le z \le 0.75$. The z=0.2 histogram (not
shown here) can also be used to define the peak.

As an independent check on the slope of the correlation in those redshift intervals where the peak distribution
can be identified, we also used the upper envelope of the distribution which we define
by the fifth largest \Ledd\ (the number of sources with larger \Ledd\ is so small in some groups
that we suspect incomplete sampling).
Obviously, the upper envelop is not affected by the flux limit of the sample.
However, the redshift bins must contain enough sources to justify this method. This may be a problem for very
high luminosity sources at low redshift because of the limited area covered by the SDSS sample.

\begin{figure}
\plotone{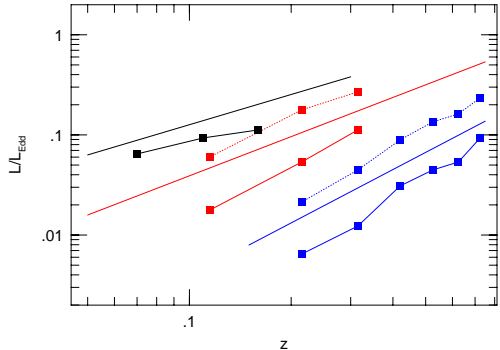}
\caption{\Ledd\ vs. z for various mass groups. The color symbols correspond to the two mass groups where
solid line represent  peaks and dotted lines represent the upper envelopes of the distributions,
as explained in the text.
 Black lines and symbols represent a third mass group with M=$10^{6.8-7.2}$ \Msun. The straight
solid lines are the results of the truncated permutation test for the same mass groups (eqn.\ref{eq:Ledd_z_1}
and Table 1).
}
\label{fig:envelope_peak}
\end{figure}

The results of the test are shown in Fig.~\ref{fig:envelope_peak}.
 We also show the peaks of the distribution for a third
mass group, with M$=10^{6.8-7.2}$ \Msun , for those redshifts where the peak of the \Ledd\  distribution
is higher than the flux limit of this group. In this mass range, this is only true for $z<0.25$ since there are
very few such low mass sources at higher redshifts because of the flux limit.
The results can be presented as simple power-laws in redshift ranging from \Ledd$\propto z^{0.8}$ for
 for M(BH)=$10^7$ \Msun\ to about \Ledd$\propto z^{2.2}$  for
M(BH)=$10^{8.6}$ \Msun. Given the small number of points in the various
redshift intervals, and the differences between the slopes obtained from the peaks and the
envelopes of the various histograms, we estimate
an uncertainty of about 0.3 on the slopes obtained with this method.

\subsubsection{The truncated permutation test}
The determination of bivariate distributions from truncated astronomical data is a well known problem. Such data sets
are very common and the flux limited sample considered here is only one such example. The problem is to test such
truncated bivariate data sets for
statistical independence without a priori knowledge of the real underlying correlation. This issue is
discussed in great detail by Efron \& Petrosian (1992; see this paper for earlier attempts to address this issue)
who suggest powerful nonparametric permutation tests that can be applied to data sets of this type. The paper describes
tests for nontruncated as well as truncated data sets and apply them to two astronomical cases that are rather
similar to the sample in question. In particular, they describe a rank-based statistics that is powerful, robust and easily
administrated to data sets like ours. 

The normalized rank statistic of Efron and Petrosian (1992), $T_i$,
is used to define a test statistic
\begin{equation}
t_w(data) = \Sigma_i w_i T_i / (\Sigma_i w_i^2)^{1/2} \,\, ,
\end{equation}
where \v{w}$=(w_1,w_2,....w_n)$ is a vector of weights. $t_w(data)$ has mean 0 and variance 1 and is distributed normally
thus allowing to obtain the best value and the confidence limits using the normal approximation (see Efron \&
Petrosian 1992 \S3 for more details and definitions). The choice made here, and recommended also in Efron \& Petrosian (1992),
assumes equal weights for all points, i.e. \v{w}$=(1,1,....1)$.
The application of the permutation test involves a guess of the real distribution which recovers the null hypothesis
of independence between the two variables in questions (in our case \Ledd\ and z).

We have used truncated permutation tests to check two families of ad hoc luminosity redshift dependences:
\begin{equation}
L/L_{Edd} \propto z^{\gamma(M)} \,\, ,
\label{eq:Ledd_z_1}
\end{equation}
and
\begin{equation}
L/L_{Edd} \propto (1+z)^{\delta(M)} \,\, .
\label{eq:Ledd_z_2}
\end{equation}
This involved changing the guessed values of $\gamma$ and $\delta$, for various mass groups, and looking for
their best values (i.e. those $\gamma$ and $\delta$ resulting in $t_w=0$) as well as the 90\% confidence limits
 (those value giving $t_w=-1.645$ and $t_w=1.645$). The mass groups used were limited to be
0.2 dex wide except for the lowest mass group, around M(BH)=$10^7$ \Msun, where the small number of objects dictates a wider range
(0.4 dex). The results are listed in Table 1 where we also specify the redshift range for each group
(the largest mass group contains no objects at small z and the lowest mass group contains very few sources 
at $z>0.25$ hence the need to define a redshift range).

The results of the test are shown in Fig.~\ref{fig:envelope_peak} for the case of eqn.~\ref{eq:Ledd_z_1}. They are
consistent with the results of the peak distribution method and lend support
to the interpretation that the peak of the distribution has indeed been recovered for the mass groups and redshift
intervals discussed above. These results are further discussed and compared with the cosmic star formation rate in \S4.2

\begin{table}
\caption{Fit coefficients for \Ledd\ vs. $z$ (eqns. 4 \& 5)}
\begin{center}
\begin{tabular}{cccc}
\hline
$\log M$ (\Msun)   & $\Delta z$   &   $\gamma$ (90\% limits)    & $\delta$ (90\% limits) \\
\hline
7.0                &  0.05--0.3   &   1.02 (0.91--1.13)         &  8.6 (7.7--9.7)       \\
7.5                & 0.05--0.75   & 1.25 (1.15--1.34)           &  7.6 (7.1--8.05)    \\
8.0                & 0.1-0.75     & 1.61 (1.54-1.67)            &  6.8 (6.6--7.1)     \\
8.5                & 0.20--0.75   & 1.79 (1.70--1.87)           &  6.1 (5.7--6.3)    \\
9.0                & 0.25--0.75    & 2.30 (2.15--2.47)           &  7.6 (7.1--8.1))  \\
\hline
\end{tabular}
\end{center}
\label{table:Ledd-z}
\end{table}

\subsection{FeII/\hb\ and iron abundance}

\begin{figure}
\plotone{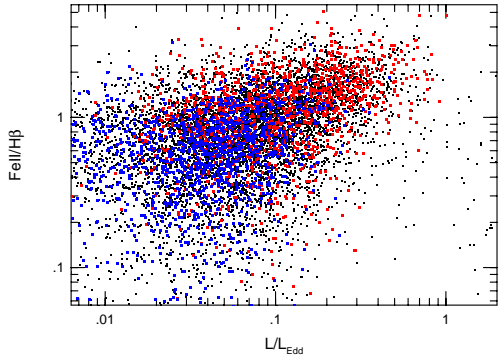}
\caption{FeII/\hb\ vs. \Ledd\ for our SDSS sample of type-I RQ-AGNs.
Symbols are the same as those of Fig.1.
}
\label{fig:feiihb_mdot}
\end{figure}

Fig.~\ref{fig:feiihb_mdot} shows the dependence of FeII/\hb\ on \Ledd\ for our sample
of type-I SDSS AGNs.
As clear from the diagram, and has been verified by standard regression analysis, 
FeII/\hb\ and \Ledd\ are
significantly correlated and larger \Ledd\ are associated with stronger
FeII and/or weaker \hb\ lines. This conclusion applies for all sources with \Ledd$\gtsim 0.03$ while
at lower accretion rates the scatter is too large to see a definite trend.
Such a correlation has been suspected in the past and has studied in 
several previous papers (see Netzer et al. 2004 and references
therein). However, it was never tested in such a large sample and over the entire 0--0.75 redshift range.

We have also studied the dependence of FeII/\hb\ on redshift in the entire sample as well as in
the various mass groups. Simple linear regression suggests that, for RQ-AGNs,
FeII/\hb$\propto z^{-0.22 \pm 0.02}$. The slope is somewhat steeper for RL sources
($-0.38 \pm 0.07)$ and for the mass group of $10^{7.7-7.8}$ \Msun\ objects ($-0.40 \pm 0.05$).
The large number of points in all those  groups of sources results in formally significant
correlations. However, detailed examination shows that the results depend strongly on the measured FeII/\hb\
at the low redshift low luminosity end of the distribution.
As explained in \S2.2, some of these cases, especially those with extremely broad lines, are difficult to measure
and are therefore uncertain because of the large stellar contribution. There are indeed differences in the mean FeII/\hb\ between
the $z<0.2$ and $z>0.5$ redshift bins but not a smooth continuous change as a function of redshift.
In view of this, we consider the evidence for redshift dependence of FeII/\hb\ as marginal.
This uncertainty does not affect the FeII/\hb\ vs. \Ledd\ correlation discussed earlier.

The modeling of FeII line formation in AGN clouds is extremely complicated.
It involves hundreds of energy levels and thousands of transitions. Earlier detailed studies by
Netzer and Wills (1983) and Wills, Netzer \& Wills (1985), that included most of the important
atomic processes, reached the conclusion that the FeII line intensities are too uncertain to be used
as iron abundance indicators. Major improvements in atomic data (Sigut \& Pradhan 2003),
combined with improved treatment of some of the line excitation processes (Baldwin et al. 2004),
resulted in a better understanding of several strong FeII features but changed little with
regard to the main conclusion: there are several mechanisms capable of enhancing
 FeII emission relative to the hydrogen Balmer lines and increased
iron abundance is only one of them .

The correlation shown in Fig.~\ref{fig:feiihb_mdot} provides a new way to solve the iron abundance
problem. As shown in S04, the N/C abundance (and  metalicity in general) as measured from the
\nv/\civ\ line ratio (Hamann et al. 2002 and references therein) is strongly
correlated with \Ledd\ over a large luminosity range.
This result is in conflict with the earlier studies by Hamann and Ferland (1993; 1999) who
suggested that the N/C ratio depends on source luminosity (see also Dietrich et al. 2002;
Dietrich et al. 2003 for the possible dependence on M(BH)). 
The S04 finding is based mostly on the high metalicity of low luminosity
narrow line Seyfert 1 galaxies (NLS1s) that do not fall on the original Hamann and Ferland (1993) 
correlation (Shemmer and Netzer 2002). The high metalicity of NLS1s gains further support by the 
detailed analysis of the absorption line spectrum of Mrk 1044 (Fields et al. 2005).

Given the dependence of metalicity on \Ledd, and the 
correlation of FeII/\hb\ with \Ledd\ shown in Fig.~\ref{fig:feiihb_mdot},
we suggest that FeII/\hb\ is an iron abundance indicator for all sources with \Ledd$\gtsim 0.03$.
 This conclusion does not rely
on uncertain atomic models and complicated radiative transfer in photoionized BLR clouds. It is based
on observed correlations in large AGN samples. There is no accurate way to calibrating the iron-to-hydrogen abundance
because of the uncertain modeling. However, we can compare this relationship to 
the S04 correlation of \nv/\civ\ vs. \Ledd\ (their Fig. 6). The
comparison shows that the slopes of the two correlations are similar. For example, the use of the BCES method with
the assumption of a uniform error of 0.1 dex in both axes gives FeII/\hb$\propto$\Ledd$^{0.7}$.
Assuming that Fe/H scales with O/H, and using the Hamann et al. (2002) calculation to calibrate N/C vs.
metalicity, we find that solar Fe/H corresponds to FeII/\hb$\simeq 0.6$.

\subsection{The growth time of massive black holes}
Next we assume that BH growth is dominated by gas accretion and BH mergers are of secondary importance.
Given this assumption, we can calculate the growth time using the Salpeter (1964) original expression as
\begin{equation}
t_{\rm grow} = 4 \times 10^8 \frac {\eta /(1- \eta) }{ L/L_{Edd}} \log \frac{M}{M_{\rm seed}}
    \frac{1}{f_{\rm active}} \,\, {\rm yr} \, ,
\label{eq:t_grow}
\end{equation}
where $f_{\rm active}$ is the fraction of time the BH is active. For simplicity we assume a mass
independent accretion efficiency of $\eta=0.2$, to reflect the assumption of a non-zero
angular velocity BHs, and a relatively large seed BH mass of $M_{seed}=10^4$ \Msun.
Obviously there is a range in both properties
(e.g. Heger and Woosley, 2002; Ohkubo et al. 2006) but the 
analysis is not very sensitive to either of those provided $M_{seed}$ is not much
larger than assumed here.

\begin{figure}
\plotone{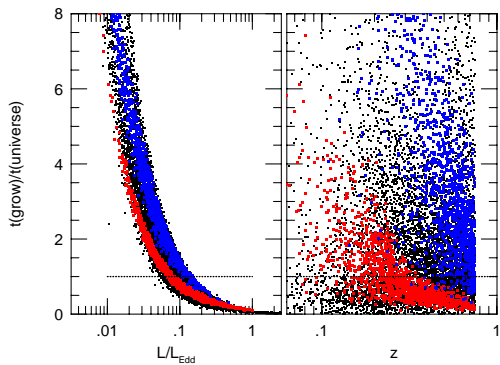}
\caption{BH growth-time, relative to the age of the
universe, as a function of \Ledd\ (left) and
redshift (right). Objects above the horizontal line at t(grow)/t(universe)=1 must
have had past episodes of faster accretion.
Symbols are the same as those of Fig.1.
}
\label{fig:growth_time}
\end{figure}

Given the above growth time, and assuming, $f_{\rm active}=1$, we can calculate for each BH the time
required to grow to its observed mass relative to the age of the universe at the observed redshift,
$t_{grow}/t_{universe}$.
This is shown in Fig.~\ref{fig:growth_time} for all sources as functions of both \Ledd\ and redshift.
Evidently, most (66\%) BHs did not have enough time to grow to their measured
mass given their observed accretion rate. The fraction of such sources found here is, of course, an upper limit,
since all BHs missing from our sample, due to their very small \Ledd\ and the SDSS flux limit, also belong in this
group of objects. Given the assumption of growth by accretion, we find that in the majority of cases, there must have been
one or more past episodes where the accretion rate was higher than the value measured here.
As shown in \S4 below, this conclusion is of major importance in understanding the
time dependent metal enrichment in AGNs.

\subsection{Redefining narrow line type-I AGNs (NLAGN1s)}

The subgroup of NLS1s is usually assumed to contain AGNs with the largest
accretion rates (e.g. Grupe et al. 2004 and references therein).
Such sources were first  discussed in Osterbrock and Pogge (1985) who introduced the
 defining criteria of FWHM(\hb)$ \le 2000$ \kms.
  NLS1s are known to have other extreme properties such as extremely
steep soft X-ray continuum and larger than average FeII/\hb.

The recent study by S04 includes H and K-band spectroscopy of about 30
high redshift, very high luminosity AGNs. In most of these sources \Ledd$>0.4$ yet FWHM(\hb)
is much larger than 2000 \kms. These authors suggested that the defining criteria for NLS1s
should be mass and luminosity dependent thus reflecting the fundamental physics affecting the line width.
This approach is different from the one taken by McLure and Dunlop (2004) who assumed the boundary
between the groups of narrow and broad line type-I AGNs is at 2000 \kms, independent of luminosity.
Given this assumption, they did not find a larger fraction of NLS1s among the highest luminosity sources in their sample.

\begin{figure}
\plotone{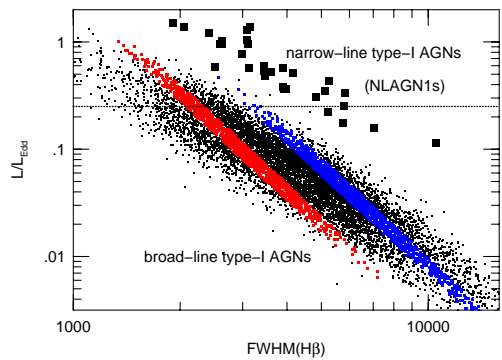}
\caption{A new definion of narrow line type-I RQ-AGN (NLAGN1s) based on accretion rate.
The large symbols represent very luminous AGNs from S04 with a median BH mass of $10^{9.5}$ \Msun.
The dividing line is drawn at \Ledd=0.25.
Symbols are the same as those of Fig.1.
}
\label{fig:fwhmhb_mdot}
\end{figure}

Our SDSS sample contains a large enough range in both M(BH) and \Ledd\ to study the S04
suggestion in more detail.
To illustrate this we plot in Fig.~\ref{fig:fwhmhb_mdot} accretion rates as a function of FWHM(\hb) for
all sources and shaw, on the same diagram,
the two mass groups introduced earlier and 31 high redshift sources from S04.
The median mass for the third group is about $3 \times 10^9$ \Msun\ and the median \Lop\ about
$10^{46.6}$ \ergs\ (an orders of magnitude larger than the most luminous $z \le 0.75$ SDSS sources).
The diagram shows also a line at \Ledd=0.25 that intersects the various subgroups at different line widths.
At small $z$, this line is consistent with the Osterbrock \& Pogge (1985) definition of NLS1s in terms of
continuum luminosity and BH mass.
We suggest that this line is a more physical measure of a general group of
 ``narrow line type-I AGN'' (hereafter NLAGN1s).
Using eqns.~\ref{eq:M_L} and 2 with $f_L=7$, and normalizing the relationship to have a boundary at
\Ledd=0.25, we get either a luminosity-based definition,
\begin{equation}
FWHM(H_{beta}) \le 1870 \left [ \frac{ L_{5100} }{ 10^{44}  \, {\rm erg\,s^{-1}} } \right ] ^{0.175} {\rm km\,s^{-1}}
\end{equation}
or a definition based on black hole mass,
\begin{equation}
FWHM(H_{beta}) \le 1700 \left [ \frac{ M }{ 10^{7}  \, M_{\odot}  } \right ] ^{0.175} {\rm km\,s^{-1}} \, .
\label{eq:new_definition_narrow}
\end{equation}

\subsection{FWHM(\oiii) as a black hole mass indicator}

The discovery of a tight correlation between black hole mass and the stellar bulge velocity
dispersion ($\sigma_*$) resulted in several attempts to investigate secondary mass
indicators related to emission line widths ($\sigma_g$). This is particularly important in those cases where
the strong non-stellar continuum prevents the measurement of $\sigma_*$. For example, FWHM(\oiii) has been
suggested as  a proxy for $\sigma_*$ in type-I AGNs (e.g. Nelson 2000; Shields et al. 2003;
Grupe \& Mathur 2004; Boroson 2003; Boroson 2005).

Boroson (2003) used a small subsample of 107 objects with $z<0.4$ from the early release SDSS data to study the
FWHM(\oiii) vs. $\sigma_*$ relationship. The FWHM(\oiii)  in this study was measured from the observed line profile
with no profile fitting. He finds a real correlation between FWHM(\oiii) and the BH mass with a flattening of the
relationship at higher luminosity and a very large scatter (a factor of $\sim 5$).
In a later publication, Boroson (2005) revisited the same problem using a larger sample (about 400 sources) covering
the same redshift interval and focusing on the systemic velocity indicated by the \oiii\ line center. He finds a systematic
blueshift of the line peak with a
magnitude which is  correlated with the line width and with \Ledd.
 An extensive discussion of the merit of such techniques, including an analysis
of a large sample (1749 object) of type-II AGNs, with a median redshift of 0.1, where both $\sigma_*$ and the width
of several narrow emission lines
have been measured, is given in Greene \& Ho (2005). According to these authors, FWHM(\oiii)/2.35 is a good proxy for
$\sigma_*$ provided the strong line core, without the commonly observed extended blue wing, is used to
define FWHM(\oiii).  Greene \& Ho further suggested that deviations
between $\sigma_*$ and FWHM(\oiii)/2.35 are related to \Ledd. Large deviation probably reflect
non-virial motions due to a strong radiation field.
The \oiii\ line luminosity has also been used as a primary bolometric luminosity indicator for type-II AGNs
(e.g. Heckman et al. 2004) whose non-stellar continuum cannot be directly observed.
As shown by Netzer et al. (2006), this is associated
with a large scatter, and perhaps also a luminosity and/or redshift dependence.

\begin{figure}
\plotone{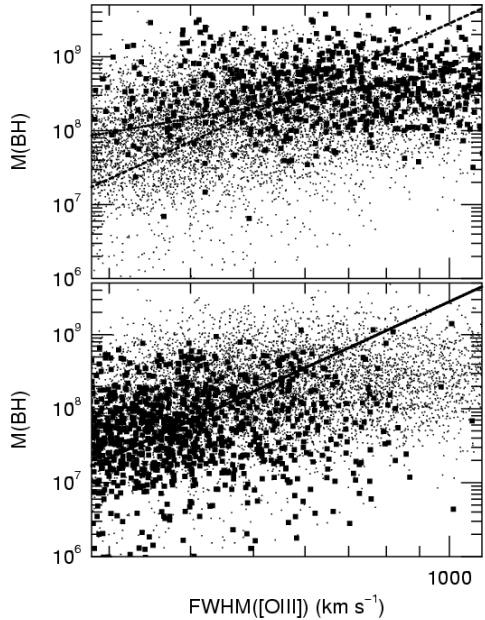}
\caption{The correlation of M(BH) vs. FWHM(\oiii) for two redshift bins:
z=0.1--0.2 (bottom) and z=0.4--0.6 (top).
The entire RQ sample is plotted as small points in both panels and the objects belonging to the
specific redshift group are plotted as large black symbols. The two lines in each panel
are best fit curves
for the redshift group in question (solid line) and the Tremaine et al. (2002) fit to the
M(BH)-$\sigma_*$ relationship (dashed line). The two fits
are basically identical for the z=0.1--0.2 group.
}
\label{fig:fwhmoiii_m}
\end{figure}

Our SDSS sample is different from all earlier ones by being much larger and by covering a wider redshift
range. It can thus be used to test the earlier findings regarding type-I AGNs at higher luminosities and
redshifts  and with a better statistics. Unlike Boroson
(2003; 2005) and Greene \& Ho (2005), we do not restrict ourselves to high S/N spectra and attempt to fit all sources with
a detectable \oiii\ line. As explained in \S2.2, our FWHM(\oiii) is based on a single Gaussian fit to the core of the
line, similar to the method of Greene \& Ho (2005).
We have  correlated these measurements with M(BH) and \Ledd\ measured here and show the results in
Fig~\ref{fig:fwhmoiii_m} as M(BH) vs. FWHM(\oiii) for the entire sample as well as for
two redshift groups: z=0.1--0.2 and z=0.4--0.6. To guide the eye, we have calculated the best
slope for each of the groups
using the BCES method assuming a uniform error of 0.1 dex and plot them
on the diagram together with a curve representing the Tremaine et al. (2002) parametrization of the
M(BH)-$\sigma_*$ relationship.

Fig.~\ref{fig:fwhmoiii_m} clearly shows the non-linear dependence of log(M(BH)) on
log(FWHM(\oiii)).
Expressing the
correlation as M(BH)$\propto$FWHM(\oiii)$^{\alpha(z)}$
we find a range of $\alpha(z)$ as listed in Table 2.
As evident from the table, the slope is decreasing with redshift and the correlation becomes
insignificant at $z>0.5$. As a curiosity, the agreement with the Tremaine et al. (2002) best slope ($\alpha=4.02$)
 is almost perfect over the redshift interval of 0.1--0.2 (see bottom panel of diagram).
We suspect that the overall agreement
between $\sigma_*$ and FWHM(\oiii)/2.35 found by Grupe \& Mathur (2004) and by Greene \& HO (2005),
is the result of those samples being dominated by low redshift sources.

\begin{table}
\caption{Fit coefficients for M(BH) vs. FWHM(\oiii) as defined in \S3.5}
\begin{center}
\begin{tabular}{cccc}
\hline
$\Delta z$ & $\alpha(z)$ & \Ledd\      & $\beta(L/L_{Edd})$ \\
\hline
0.05--0.15 & $5.8\pm1.5$ & 0.01--0.03  & $3.4\pm0.1$  \\
0.10--0.20 & $4.0\pm0.6$ & 0.10--0.15  & $2.7\pm0.7$  \\
0.37--0.43 & $1.6\pm0.2$ & 0.30--0.50  & $2.1\pm0.2$  \\
0.66--0.75 & $1.5\pm0.9$ & all         & $2.9\pm0.4$  \\
\hline
\end{tabular}
\label{table:m_oiii}
\end{center}
\end{table}

High accretion rate can also influence
the \oiii\ line width by introducing a non-virial component to the line profile, perhaps due to
acceleration by radiation pressure force. This has been studied by Netzer et al. (2004), Boroson (2005) and
Greene \& Ho (2005).
 Netzer et al. found that FWHM(\oiii) correlates with \Lop, M(BH) and \Ledd\ but the
dependence on \Ledd\ is weaker than the dependences on the luminosity and the mass. To check this in the present case, we
 divided the SDSS sample into groups according to \Ledd\ and assumed that
M(BH)$\propto$FWHM(\oiii)$^{\beta(L/L_{Edd})}$.
We find a clear decrease of $\beta$ with \Ledd. The computed values
are given in Table 2. Thus, larger \Ledd\ results in a flatter
M(BH)-FWHM(\oiii) relationship.

The redshift dependence of the M(BH)-FWHM(\oiii) relationship can be explained in several ways.
First, selection effects must be important as more and more low mass, low luminosity and
small \Ledd\  sources drop from the sample at higher redshifts.
We suspect that the paucity of small mass black holes at high redshifts cannot explain the entire change as the overall range
in M(BH) at $z>0.5$ is quite large (about 1.3 dex). The increasing number of high \Ledd\ sources
is probably more important as this tends to flatten the relationship. This is in line with the  Greene \& Ho
(2005) suggestion and was tested quantitatively by using their recipe (their eqn. 1)
 to correct the observed FWHM(\oiii) for
\Ledd. The procedure results in a somewhat better agreement with the Tremaine
et al. (2002) relationship over part (but not all) of the range in FWHM(\oiii). This by itself
does not explain the observed spread. We conclude that on top of a large intrinsic spread in
all quantities related to the \oiii\ line, there are also systematic effects that limit, severely,
the usefulness of this approach.
Thus, the methods used to obtain bolometric luminosity and BH mass from the luminosity and the width
of the \oiii\ line
suffer from a large scatter and, perhaps, systematic uncertainties.

An interesting idea is a real redshift
evolution of the M(BH)-$\sigma_*$ relationship. In this case, FWHM(\oiii)/2.35 can
be used as a surrogate for $\sigma_*$ at small z yet the non-parallel evolution of the bulge and the BH mass
requires a modification at larger redshifts.
Further discussion of this idea is beyond the scope of the paper.

\subsection{Other correlations}

The SDSS sample is an invaluable source for studying several other correlations of line
and continuum properties in type-I AGNs. Such a study goes beyond the scope of the present
work that focuses on mass, accretion rate and metalicity as a function of redshift. Since most those
correlations can easily be computed, we
list them below, with only a brief discussion and leave the more detailed investigation
to a forthcoming publication.

Table~3 shows the results of our Spearman rank correlation analysis
for several line and continuum properties. 
Significant correlations, with a chance probability smaller
than $10^{-5}$, are marked by either a $+$ or a
$-$, depending on the slope of the correlation.
In particular, we
find a negative Baldwin relationship (Baldwin 1977) for both \hb\ and the FeII lines. The former has already
been found by Croom et al. (2004) in their study of the  2dF AGN sample.
We also find a tendency for the fractional luminosity of the red part of the \hb\ line 
(measured relative to the position of
\oiii) to increase with continuum luminosity. This is an opposite trend to the well known blue
shift of the \civ\ line in high luminosity sources and does not seem to be associated
with \Ledd. We note that the definition of the systemic velocity is a bit ambiguous since we rely
on the measured wavelength of the \oiii\ line. Yet, this lines was shown by Boroson (2005) to have
a noticeable blueshift in large \Ledd\ AGNs.

\begin{table}
\caption{Correlations with a chance probability smaller than $10^{-5}$}
\begin{center}
\begin{tabular}{llccc}
\hline
Property A & Property B & RQ-AGNs  & RL-AGNs  & M(BH)=10$^{7.5-7.8}$ \Msun\ \\
\hline
EW(\hb)    & \Lop\           & +   &    &   \\
EW(\hb)    & M(BH)           & +   & +  &  \\
EW(\hb)    & \Ledd\          & $-$ & $-$&  \\
EW(FeII)  & \Lop\            & +   &    & + \\
EW(FeII)  & M(BH)            & $-$ & $-$&   \\
EW(FeII)  & \Ledd\           & +   & +  & + \\
FeII/\hb\   & \Lop\          & $-$ &    & + \\
FeII/\hb\   & M(BH)          & $-$ &$-$ &   \\
FeII/\hb\   & \Ledd\         & +   & +  & + \\
FeII/\hb\   & EW(\hb)        & $-$ & $-$& $-$ \\
\hb(red)/\hb(total)& \Lop\   & +   & +  & +  \\
\hb(red)/\hb(total)& M(BH)   & +   & +  &   \\
\hb(red)/\hb(total)& EW(\hb) & +   & +  & +  \\
\hb(red)/\hb(total)&FeII/\hb\ &$-$  & $-$& $-$ \\
\hline
\label{corr_table}
\end{tabular}
\label{table:correlations}
\end{center}
\end{table}

\section{Discussion}

The main new results presented here are the measurement of the accretion rate, in several mass groups, as a function of
redshift and the correlation between black hole growth rate and BLR metalicity.

\subsection{Black hole growth and BLR metalicity}

The new results concerning the iron abundance and the growth time can be combined to infer
the changes in the metal abundance of the BLR gas as a function of time. Since FeII/\hb\ is an iron abundance
indicator which depends on \Ledd, and since most type-I AGNs had
higher accretion rate episodes in their past, we cannot avoid the conclusion that those past episodes
were characterized by enhanced metalicity relative to the one measured here. Computing BH growth times
for the 31 high luminosity sources in S04, and noting the dependence of N/C on \Ledd\ found by those authors,
 results in a similar conclusion, despite of the much larger redshifts
of these sources.
Thus, BLR metalicity can go through cycles and  is not monotonously decreasing with decreasing
redshift.

The emerging picture of a complex metalicity, accretion rate and cosmic time dependences is not easy to explain.
One possibility is that episodes of fast accretion onto the central BH are associated
 with enhanced, large scale star formation episodes that
  lead  to increased metalicity throughout the host galaxy or at least in its central kpc. If the enriched material
finds its way to the center of the system, increasing both \Ledd\ and the metal content of the BLR, it
would explain all observed properties.
Such a scenario must also involve ejection and/or accretion of the enriched gas toward the end of the active phase
such that the metal content of the BLR at later times, when the fast accretion phase is over,
is reduced relative to the peak activity phase. The formation of enriched BLR clouds from the wind of
an enriched accretion disk is just one of several such possibilities.

The (somewhat uncertain) dependence of
FeII/\hb\ on redshift suggests that the general, time averaged BLR iron abundance is slowly increasing
with cosmic time
while undergoing much larger fluctuations associated with the fast accretion rate episodes.
Finally, if the time scale of channeling such enriched starburst gas into the center is short compared with
one episode of BH growth, we would expect to see strong starburst activity associated with the
BH activity in many AGNs.

\subsection{Black hole growth and cosmic star formation}

Our new results clearly show the changes in BH mass and accretion rate with redshift. For example,
the histograms shown in Fig.~\ref{fig:histogram_M4_M5_M6} clearly illustrate that the smaller BHs are the fastest
accretors at all redshifts studied here although large mass BHs at high z can accrete as fast as small mass BHs at low z.

Given the results shown in Fig.~\ref{fig:envelope_peak} and listed in Table 1, we can compare the BH accretion rates found here
 and the known star formation rate over the redshift interval of 0--0.75.
 For this we use eqns.~\ref{eq:Ledd_z_1} and \ref{eq:Ledd_z_2} and various publications describing evolutionary models and
star formation rate measurements (e.g. Chary \& Elbaz 2001 and references therein).
A simple fit to some of those results gives
$\rho_* \propto z^{\gamma}$
or
$\rho_* \propto (1+z)^{\delta}$
where $1.4 < \gamma < 2$ and $6< \delta < 9$. The large range of slopes represents the 
uncertainties on the data as well as on the various evolutionary models (see
Chary \& Elbaz 2001 for more information).
This is a universal rate averaged over all
starburst galaxies of all morphologies and luminosities. The range of slopes is somewhat narrower than the
one found here for BH growth rates (Table 1) and the middle of the range is in rough agreement with the
growth rate of M=$10^8$ \Msun\ BHs.
Thus, the average BH growth rate for $z \le 0.75$ AGNs seems to agree with the star formation rate
over the same redshift interval.

Marconi et al. (2004) used a detailed modeling of several observed AGN luminosity functions to argue
that accretion onto BHs, integrated over the entire
AGN population, proceeds at a rate similar to the star formation rate at small redshifts.
Their model fitting to the data over the redshift range 0--0.6, assuming \Ledd=1,
 can be described by
 $ L({\rm all~AGN)} \propto (1+z)^{4.2}$  (A. Marconi, private communication).
 Obviously, the Marconi et al. (2004) results cannot be directly compared with ours. Their total accretion rate
 (i.e. luminosity) was calculated for the entire AGN population and the redshift dependence was obtained under
 the assumption of a uniform \Ledd.

The new results shown here go one step further by neglecting the uniform \Ledd\ assumption and
by demonstrating that different mass groups are characterized by different growth rates.
A more detailed comparison between the two studies requires an additional information about sources
that are missing from the SDSS data but are found in X-ray selected samples.
This is beyond the scope of the present work.

\begin{acknowledgements}

Funding for this work has been
provided by the Israel Science Foundation grant 232/03 and by the Jack Adler chair of Extragalactic astronomy
at Tel Aviv University. We are grateful to A. Marconi for useful information and discussion. We also thank
V. Petrosian for introducing us to truncated permutation tests.
Useful comments by an anonymous referee helped to improved the presentation of the paper.
HN acknowledges an Humboldt foundation prize and thank the host institution, MPE Garching, 
where much of this work has been conducted. HN also thanks TIFR in Mumbai for hospitality and support
during the final stages of this work.

Funding for the creation and distribution of the SDSS Archive has been
provided by the Alfred P. Sloan Foundation, the Participating
Institutions, the National Aeronautics and Space Administration, the
National Science Foundation, the U.S. Department of Energy, the
Japanese Monbukagakusho, and the Max Planck Society.
The SDSS Web site is http://www.sdss.org/.
The SDSS is managed by the Astrophysical Research Consortium (ARC) for
the Participating Institutions. The Participating Institutions are The
University of Chicago, Fermilab, the Institute for Advanced Study, the
Japan Participation Group, The Johns Hopkins University, Los Alamos
National Laboratory, the Max-Planck-Institute for Astronomy (MPIA),
the Max-Planck-Institute for Astrophysics (MPA), New Mexico State
University, University of Pittsburgh, Princeton University, the United
States Naval Observatory, and the University of Washington.
This research has made use of the NED database which is operated by
the Jet Propulsion Laboratory, California Institute of Technology,
under contract with the National Aeronautics and Space Administration.

\end{acknowledgements}

\end{document}